\documentclass[a4paper,fleqn,usenatbib]{mnras}

\usepackage{newtxtext,newtxmath}

\usepackage[T1]{fontenc}
\usepackage{ae,aecompl}

\usepackage{graphicx}	
\usepackage{amsmath}	
\usepackage{longtable}
\usepackage{pdflscape}

\title[SN~Ia snapshot distances]{The snapshot distance method: estimating the distance to a Type Ia supernova from minimal observations}

\author[B. E. Stahl et al.]{Benjamin E. Stahl,$^{1,2}$\thanks{E-mail: benjamin\_stahl@berkeley.edu}\thanks{Marc J. Staley Graduate Fellow.}
Thomas de Jaeger,$^{3,1}$\thanks{Bengier Postdoctoral Fellow.}
WeiKang Zheng,$^{1}$
and Alexei V. Filippenko$^{1,4}$\thanks{Miller Senior Fellow.}
\\
$^{1}$Department of Astronomy, University of California, Berkeley, CA 94720-3411, USA\\
$^{2}$Department of Physics, University of California, Berkeley, CA 94720-7300, USA\\
$^{3}$ Institute for Astronomy, University of Hawaii, 2680 Woodlawn Drive, Honolulu, HI 96822, USA\\
$^{4}$Miller Institute for Basic Research in Science, University of California, Berkeley, CA 94720, USA
}

\date{Accepted XXX. Received YYY; in original form ZZZ}

\pubyear{2021}

\begin{document}
\label{firstpage}
\pagerange{\pageref{firstpage}--\pageref{lastpage}}
\maketitle

\begin{abstract}
We present the snapshot distance method (SDM), a modern incarnation of a proposed technique for estimating the distance to a Type Ia supernova (SN~Ia) from minimal observations. Our method, which has become possible owing to recent work in the application of deep learning to SN~Ia spectra (we use the \texttt{deepSIP} package), allows us to estimate the distance to an SN~Ia from a single optical spectrum and epoch of $2+$ passband photometry --- one night's worth of observations (though contemporaneity is not a requirement). Using a compilation of well-observed SNe~Ia, we generate snapshot distances across a wide range of spectral and photometric phases, light-curve shapes, photometric passband combinations, and spectrum signal-to-noise ratios. By comparing these estimates to the corresponding distances derived from fitting all available photometry for each object, we demonstrate that our method is robust to the relative temporal sampling of the provided spectroscopic and photometric information, and to a broad range of light-curve shapes that lie within the domain of standard width-luminosity relations. Indeed, the median residual (and asymmetric scatter) between SDM distances derived from two-passband photometry and conventional light-curve-derived distances that utilise all available photometry is $0.013_{-0.143}^{+0.154}$\,mag. Moreover, we find that the time of maximum brightness and light-curve shape (both of which are spectroscopically derived in our method) are only minimally responsible for the observed scatter. In a companion paper, we apply the SDM to a large number of sparsely observed SNe~Ia as part of a cosmological study.
\end{abstract}




\section{Introduction}
\label{sec:introduction}

Type Ia supernovae (SNe~Ia) result from the thermonuclear runaway explosions of carbon/oxygen white dwarfs in binary star systems \citep[e.g.,][]{Hoyle1960,Colgate1969,Nomoto1984} in which the stellar companion may \citep[e.g.,][]{Webbink1984,IbenDD} or may not \citep[e.g.,][]{WhelanSD} be another white dwarf. Despite our incomplete understanding of SN~Ia progenitor systems and explosion mechanisms \citep[see][for a recent review]{JhaReview}, it remains an empirical fact that SNe~Ia (or at least, a subset thereof) follow photometric \citep[e.g.,][]{Phillips1993,Riess1996,mlcs2k2} and spectroscopic \citep[e.g.,][]{Nugent1995} sequences with regard to peak luminosity. This fact, in conjunction with their extraordinary luminosities, makes SNe~Ia immensely valuable as cosmological distance indicators. Indeed, exploitation of the aforementioned photometric sequence, whereby the width of an SN~Ia light curve is used to standardise its peak luminosity (hence the ``width-luminosity relation'' moniker), along with photometrically-derived corrections for reddening due to host-galaxy dust, led to the discovery of the accelerating expansion of the Universe \citep{Riess1998,Perlmutter1999}.

As the photometric samples of nearby \citep[redshift $z \lesssim 0.1$;][]{CfA1,CfA2,CfA3,Ganeshalingam2010,CSP1,CSP2,CSP3,Foundation,S19} and distant \citep[$z \gtrsim 0.1$; e.g.,][]{ESSENCE1,SDSSSNe,ESSENCE2} SNe~Ia have grown, parameterisations of the SN~Ia width-luminosity relation (WLR) have become increasingly robust \citep[e.g.,][]{SALT2,SNooPy}. Together, these have aided in placing increasingly stringent constraints on the composition \citep{ESSENCECosmo,KesslerCosmo,SNLSCosmo1,SNLSCosmo2,Suzuki2012,G12,Betoule2014,Scolnic2018} and present expansion rate \citep{Riess2016,Riess2019} of the Universe.

At the same time, the spectroscopic sample of SNe~Ia has grown considerably \citep[e.g.,][]{bsnipI,Blondin2012,Folatelli2013,S20}. Consequently, there has been forward progress in identifying spectroscopic parameters to potentially improve the precision of SN~Ia distance measurements \citep[e.g.,][]{Bailey2009,Wang2009,Blondin2011,bsnipIII,Fakhouri2015,Zheng_empirical,kaepora,SUGAR}. Relatedly, recent work has demonstrated that $\Delta m_{15}$, a measure of light-curve shape --- and hence, of peak luminosity via the SN~Ia WLR --- can be recovered from a single optical spectrum with a high degree of precision through the use of convolutional neural networks \citep[using, e.g., the \texttt{deepSIP}\footnote{\url{https://github.com/benstahl92/deepSIP}} package;][S20 hereafter]{deepSIP}. Moreover, owing to the data-augmentation strategy employed in the training of its models, \texttt{deepSIP} is robust to the signal-to-noise ratios (SNRs) of spectra it processes (we defer the reader to S20 for more details). In addition to $\Delta m_{15}$, \texttt{deepSIP} can also, again from a single optical spectrum, predict the the time elapsed since maximum light --- i.e., the phase --- of an SN~Ia in the rest frame, from which the time of maximum brightness, $t_\mathrm{max}$, can be calculated. Together, these two quantities ($\Delta m_{15}$ and $t_\mathrm{max}$) amount to half of those that are conventionally derived from a light-curve-fitting analysis, with other two being (i) a measure of the extinction produced by dust in the SN's host galaxy and (ii) the distance to the SN \citep[see, e.g.,][for WLR implementations that function in this way]{mlcs2k2,SNooPy}.

As a result, a single SN~Ia spectrum --- via \texttt{deepSIP} --- can powerfully constrain the family of light curves that could \emph{possibly} correspond to that object. This motivates us to revisit the notion of a ``snapshot'' distance \citep[hereafter R98]{RiessSnapshot}: the idea that a single night's worth of SN~Ia observations --- an optical spectrum and one epoch of multiband photometry --- is sufficient to estimate the distance to an SN~Ia. Although photometric classification schemes are now available \citep[e.g.,][]{Richards-photoclassify,RAPID}, the results are not yet --- and may never be --- competitive with spectra. Hence, spectra are still the preferred method for classifying SNe \citep[see, e.g.,][for reviews of SN classification]{Filippenko1997,GY17}, and as a result, a viable method of snapshot distances could render some cosmologically-motivated follow-up photometry unnecessary, thereby conserving valuable and limited observing resources.

In this paper, we present the snapshot distance method (SDM), a modern version of the initial concept established by R98. We describe the method itself in Section~\ref{sec:method} before undertaking a rigorous and comprehensive study of its efficacy in Section~\ref{sec:sim}. We conclude with a discussion of possible variations of the SDM and anticipated uses in Section~\ref{sec:discussion}.

\section{The Snapshot Distance Method}
\label{sec:method}

\begin{figure}
 \includegraphics[width=\columnwidth]{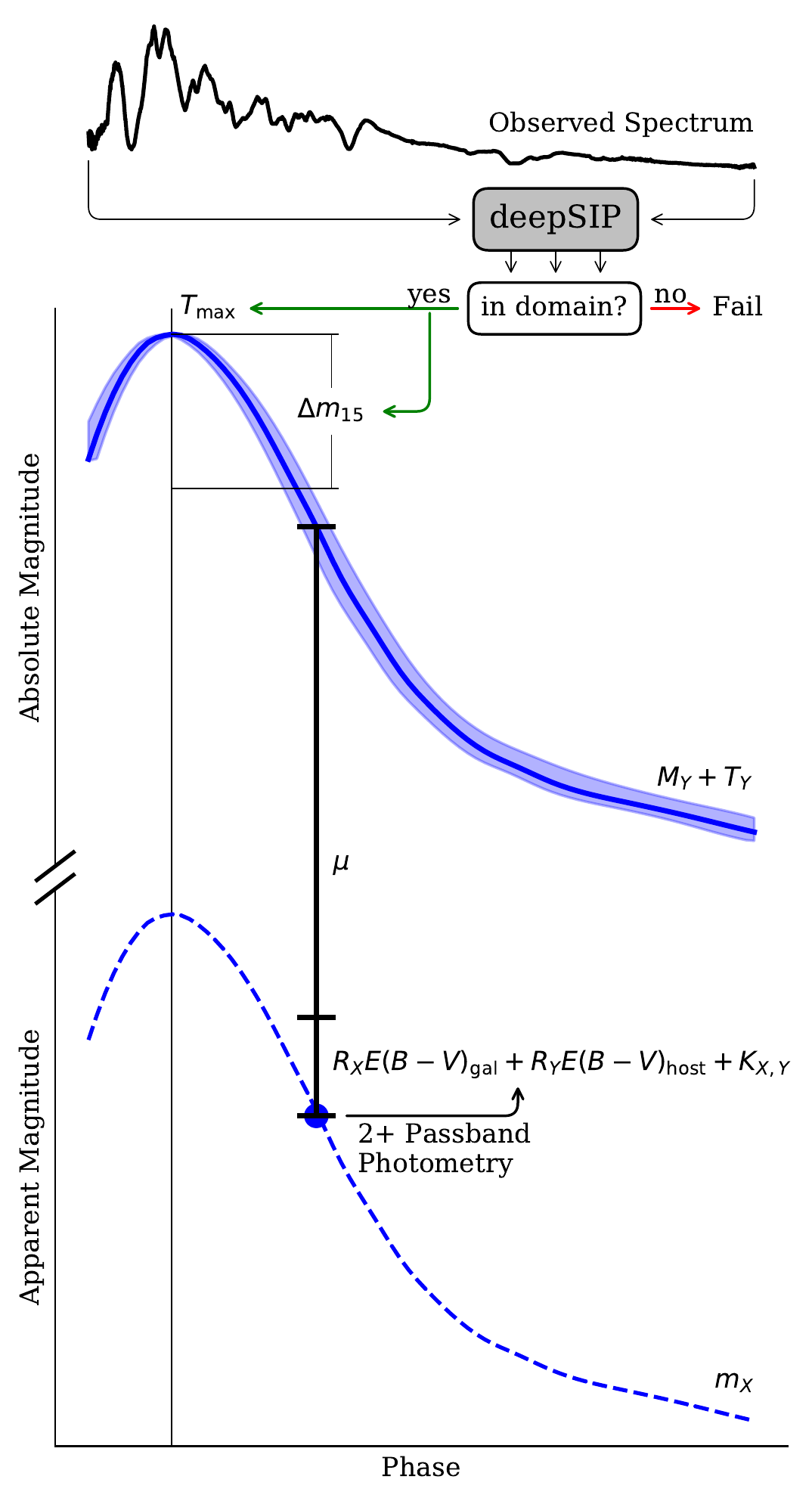}
 \caption[Schematic representation of the snapshot distance method]{Schematic representation of the SDM applied to observations of SN~2017erp \citep{S19,S20}. Using \texttt{deepSIP}, the phase and light-curve shape (each with uncertainties) are extracted from an optical spectrum. These parameters alone are sufficient to derive the intrinsic luminosity evolution, $M_Y + T_Y$, in rest-frame passband $Y$ (see Equation~\ref{eqn:ebv} for details). By comparing an observed magnitude, $m_X$, in the corresponding observer-frame passband (blue circle) to this evolution sampled at the epoch of the observed magnitude, the distance modulus can be readily derived after computing $K$-corrections and accounting for Galactic and host-galaxy reddening using a second observed magnitude in a distinct passband.\label{fig:schematic}}
\end{figure}

As demonstrated by S20, the phase and light-curve shape of an SN~Ia can be inferred (with an expected precision of $\sim 1.0$\,d and $\sim 0.07$\,mag, respectively; see S20 for additional details) from an optical spectrum using \texttt{deepSIP}. This information, in conjunction with an apparent magnitude and an estimate of the extinction produced by host-galaxy dust (which can be derived from a single epoch of multiband photometry), is sufficient to estimate the distance to an SN~Ia. Figure~\ref{fig:schematic} provides a schematic representation of the procedure, which is also described comprehensively below.

While R98 treat this and the associated uncertainty estimation analytically within the multicolour light-curve shape (MLCS) formalism of \citet{Riess1996}, we use the spectroscopically recovered parameters (and their uncertainties) as priors in a Markov Chain Monte Carlo (MCMC) fit of the $E(B - V)$ model from the \texttt{SNooPy} light-curve fitter to the available photometry \citep[see][for details on \texttt{SNooPy} and its capabilities]{SNooPy}. This has the advantage of allowing for the best estimates of the time of maximum brightness and light-curve shape --- which are derived solely from an optical spectrum --- to be updated in light of additional evidence: the multiband photometry. The resulting distance estimate is therefore derived from parameters that extract maximal utility from the available data.

Our algorithm for estimating the distance to an SN~Ia from an optical spectrum and an epoch of multiband photometry that can, but need not, be contemporaneous, is as follows. 
\begin{enumerate}
\item Using \texttt{deepSIP}'s Model~I --- a binary classifier --- we determine if the spectrum belongs to an SN~Ia with a phase and light-curve shape\footnote{We note that $\Delta m_{15}$ is a \emph{generalised} light-curve-shape parameter, distinct from the traditional $\Delta m_{15}(B)$. The two may deviate randomly and systematically \citep[see Section 3.4.2 in][]{SNooPy}.} satisfying the conditions $-10 \leq \mathrm{phase} < 18$\,d and $0.85 \leq \Delta m_{15} < 1.55$\,mag, corresponding to the bounds within which \texttt{deepSIP} can reliably make continuous predictions with its other models. If the spectrum is classified as being within this ``domain'' in phase$-\Delta m_{15}$ space, we measure its phase and light-curve shape using Models II and III from \texttt{deepSIP}. The time of maximum brightness, $t_\mathrm{max}$, is then computed as the difference between the time at which the spectrum was observed minus the reconstructed phase, multiplied by a factor of $1 + z$ to express the time interval in the \emph{observer} frame. As shown in Equation~\ref{eqn:ebv} and Figure~\ref{fig:schematic}, these two parameters are sufficient to reconstruct the intrinsic luminosity evolution of an SN~Ia through the use of a WLR. In the work described herein, we use the \texttt{SNooPy} $E(B - V)$ model \citep{SNooPy}, but others could conceivably be used if \texttt{deepSIP} were retrained to predict the required light-curve-shape parameter.
\item We then perform an initial, nonlinear least-squares fit --- holding $t_\mathrm{max}$ and $\Delta m_{15}$ fixed at their \texttt{deepSIP}-determined values --- to the available photometry using the $E(B - V)$ model, which takes the mathematical form
\begin{align}
	\label{eqn:ebv}
	m_X(t - t_\mathrm{max}) = \; &M_Y(\Delta m_{15}) + T_Y\left(t_\mathrm{rel}, \Delta m_{15}\right) + \mu  + \nonumber \\
	&R_X E(B - V)_\mathrm{gal} + R_Y E(B - V)_\mathrm{host} + K_{X,Y} 
\end{align}
where $X$ ($Y$) refers to the observed (rest-frame) passband, $m$ is the observed magnitude, $t_\mathrm{rel} = (t^\prime - t_\mathrm{max})/(1 + z)$ is the rest-frame phase, $M$ is the rest-frame absolute magnitude of the SN, $T$ is a light-curve template\footnote{$M_Y(\Delta m_{15}) + T_Y\left(t_\mathrm{rel}, \Delta m_{15}\right)$ gives the absolute magnitude of an SN~Ia having the specified light-curve shape in passband $Y$ at the given phase.} generated from the prescription of \citet{Prieto2006}, $\mu$ is the distance modulus, $E(B - V)_\mathrm{gal}$ and $E(B - V)_\mathrm{host}$ are the reddening due to the Galactic foreground and host galaxy (respectively), $R$ is the total-to-selective absorption, and $K_{X,Y}$ is the $K$-correction \citep{OS68-Kcorr,Hamuy93-Kcorr,Kim96-Kcorr}. In effect, this reduces the number of parameters fit from four [$\mu, E(B - V)_\mathrm{host}, t_\mathrm{max}, \Delta m_{15}$] to two [$\mu$ and $E(B - V)_\mathrm{host}$], because the other two (i.e., $t_\mathrm{max}$ and $\Delta m_{15}$) are constrained directly by \texttt{deepSIP}. We show $K_{X,Y}$ --- which is computed by warping the appropriate SED template from \citet{Hsiao} such that performing synthetic photometry on it yields colours that match those from the observed photometry --- without its redshift, temporal, and extinction dependences for clarity. Thus, $K_{X,Y}$ depends \emph{mostly} on the supplied photometric information, but the spectroscopically derived value for $t_\mathrm{max}$ factors into the calculation of $t_\mathrm{rel}$ as shown above. Note that in the low-redshift limit (within which we primarily work herein), $X$ and $Y$ are very nearly the same passband and the $K$-corrections are small. 
\item The results from this initial fit serve as the starting point for the MCMC chains in the final fit, during which we fit for the full four parameters of the $E(B - V)$ model. In doing so, we employ Gaussian priors for $t_\mathrm{max}$ and $\Delta m_{15}$ with means (standard deviations) set to the predictions (predicted uncertainties) derived from \texttt{deepSIP}'s Models II and III, respectively. All other facets of the model --- e.g., priors for the other fitted parameters and values for static paremeters --- are left at \texttt{SNooPy} defaults \citep[see][for more details]{SNooPy}. We adopt the distance modulus resulting from this final fit as our best estimate of the SN's distance.
\end{enumerate}

\section{Validating the Snapshot Distance Method}
\label{sec:sim}

As argued in Section~\ref{sec:method}, it is possible --- in principle --- to estimate the distance to an SN~Ia from a single epoch of multiband photometry and an optical spectrum. However, before such estimates can be made with any confidence, the SDM must be subjected to a rigorous assessment to quantify its effectiveness and reliability. We endeavor to administer such a ``stress test'' by constructing snapshot distances from a masked collection of (photometrically) well-monitored objects having at least one available optical spectrum. In the following subsections we describe this collection of photometric and spectroscopic observations, the details of our validation exercise, and quantitative statements that our results substantiate.

\subsection{Data}
\label{ssec:data}

In developing \texttt{deepSIP}, S20 assembled a significant compilation of low-redshift SN~Ia optical spectra from the data releases of the Berkeley SuperNova Ia Program \citep[BSNIP;][]{bsnipI,S20}, the Harvard-Smithsonian Center for Astrophysics \citep[CfA;][]{Blondin2012}, and the Carnegie Supernova Program \citep[CSP;][]{Folatelli2013} that they then coupled to photometrically derived quantities (i.e., $t_\mathrm{max}$ and $\Delta m_{15}$) obtained by either (i) refitting the SN~Ia light curves published by the same groups \citep[][the first for the initial Berkeley sample and the last three for the CfA sample]{Ganeshalingam2010,CfA1,CfA2,CfA3}, or (ii) taking the (identically derived) fitted parameters as directly published \citep[][the former for the CSP sample and the latter for the latest Berkeley sample]{CSP3,S19}.

Altogether, this sample is nearly ideal for our purposes --- it consists of optical SN~Ia spectra spanning a wide range of phases and light-curve shapes, both of which are ultimately determined from fits to well-sampled light curves --- but we must impose two cuts on the full sample (i.e., the ``in-domain'' sample from S20 satisfying $-10 \leq \mathrm{phase} < 18$\,d and $0.85 \leq \Delta m_{15} < 1.55$\,mag; we defer the reader to S20 for more details) in order to proceed. First, we drop all spectra that were used to train\footnote{S20 allocated $\sim 80\%$ of their compilation for training, leaving the remainder (which we use in this work) for validation and testing.} \texttt{deepSIP}. This ensures that \texttt{deepSIP}-based phase and $\Delta m_{15}$ predictions used in constructing snapshot distances during validation are not unrealistically accurate. Second, we drop all spectra corresponding to SNe~Ia whose photometrically-derived parameters (e.g., $\Delta m_{15}$) were derived from the $uBVgriYJH$ observations published by \citet{CSP3} because, for simplicity, we prefer to use a consistent photometric system and set of passbands (i.e., standard $BVRI$) in the analysis described herein. Moreover, this second cut mitigates the potential for performance indicators that are favourably biased due to the fact that \texttt{SNooPy} was developed for use with CSP (and more generally, natural system) photometry --- by removing these, our analysis proceeds with \emph{only} Landolt-system data, thereby ensuring uniformity in the inputs to \texttt{SNooPy}. In the end, this leaves us with 190 spectra of 97 distinct SNe~Ia, which are collectively covered by 2450 epochs of multipassband photometry.

\subsection{Validation Strategy}
\label{ssec:simulation}

With the aforementioned dataset, we are able to test the SDM at scale. We do so by generating snapshot distances --- whereby we provide one epoch of photometry and one optical spectrum corresponding to the same SN~Ia and generate a distance estimate according to the algorithm detailed in Section~\ref{sec:method} --- exhaustively across our dataset. Our strategy is organised as follows. For each spectrum in our dataset, we generate a distinct distance estimate by providing the \texttt{deepSIP}-inferred $t_\mathrm{max}$ and $\Delta m_{15}$ values from the spectrum and every possible combination of a single photometric epoch in at least two passbands from the available photometry of the relevant SN~Ia. Thus, for the typical $BVRI$ coverage available in the photometric component of our dataset, there are 11 unique passband combinations\footnote{The 11 possible combinations for selecting $2+$ passbands from $BVRI$ are $BV$, $BR$, $BI$, $VR$, $VI$, $RI$, $BVR$, $BVI$, $BRI$, $VRI$, and $BVRI$.} and hence as many distinct distance estimates \emph{per} single epoch of photometry. Altogether, then, a total of 34,721 distinct distance estimates are attempted after we remove those photometric epochs that have rest-frame phases outside the range spanned by $-10$ and $70$\,d (i.e., the full temporal extent of the light-curve templates used in fitting), as determined relative to the \texttt{deepSIP}-inferred $t_\mathrm{max}$.

\subsection{Results}

Of the 34,721 attempted distance estimates, only 238 fail during the preliminary least-squares fitting and a further 101 fail during the final MCMC fit. In contrast, when we repeat the exercise but do not provide the spectroscopically derived parameters (i.e., we attempt to fit only the sparse photometry), over 24,000 failures occur. This is, of course, expected because \texttt{SNooPy} (and indeed, all light-curve-fitting methods) is not intended to be used with a single epoch of photometry. Though the SDM failures represent only a small proportion of all attempts, it is important to understand their origin. Our investigations reveal that the dominant mechanism in these failure modes is the phase of the supplied photometric epoch. Aggregating over the 339 total failures, the median photometric phase is $\sim 65$\,d, but just $\sim 15$\,d for those with successful fits. We reiterate that we impose an upper limit of 70\,d with respect to photometric phases that we even attempt to fit, so the fact that the median failure has a photometric phase so near to the upper bound illuminates how skewed the distribution of failures is toward late phases. More concretely, it is negligible up to $\sim 30$\,d, slowly grows from there until $\sim 60$\,d, and then blows up beyond. This follows our expectation --- SN~Ia photometric evolution progresses more slowly at later phases, and thus, it is reassuring to find that when our method breaks down, it coincides with this late-time behaviour.

Before we delve into the quantitative efficacy measures discussed in the following paragraphs, it is important that we establish clear criteria for evaluation. We will henceforth consider the parameters derived from an $E(B - V)$ model fit to all available photometry for a given well-sampled object in our data compilation to comprise a set of ``reference'' values (hereafter, \texttt{SNooPy} reference values). The goal of the SDM is therefore to reproduce the reference values of our selected light-curve fitter (i.e., \texttt{SNooPy}) using a severely limited amount of data, and although we have assumed a specific light-curve fitter in our current implementation, our algorithm is sufficiently general to transcend it --- the basic requirement would be to retrain \texttt{deepSIP} to predict the specific light-curve-shape parameter of relevance. In this spirit, we focus our subsequent study on our current SDM implementation's ability to reconstruct the \texttt{SNooPy} reference values described above, quantified in most cases by distance-modulus residuals,
\begin{align}
	\mu^\mathrm{resid} \equiv \mu^\mathrm{SDM} - \mu^\mathrm{ref},
\end{align}
where $\mu^\mathrm{SDM}$ is the distance modulus produced by the SDM for the data subset under consideration and $\mu^\mathrm{ref}$ is that produced by a standard light-curve fitter (e.g., \texttt{SNooPy} in this case) without the use of spectral information, but with the object's full light curve.

\begin{figure}
\centering
\includegraphics[width=\columnwidth]{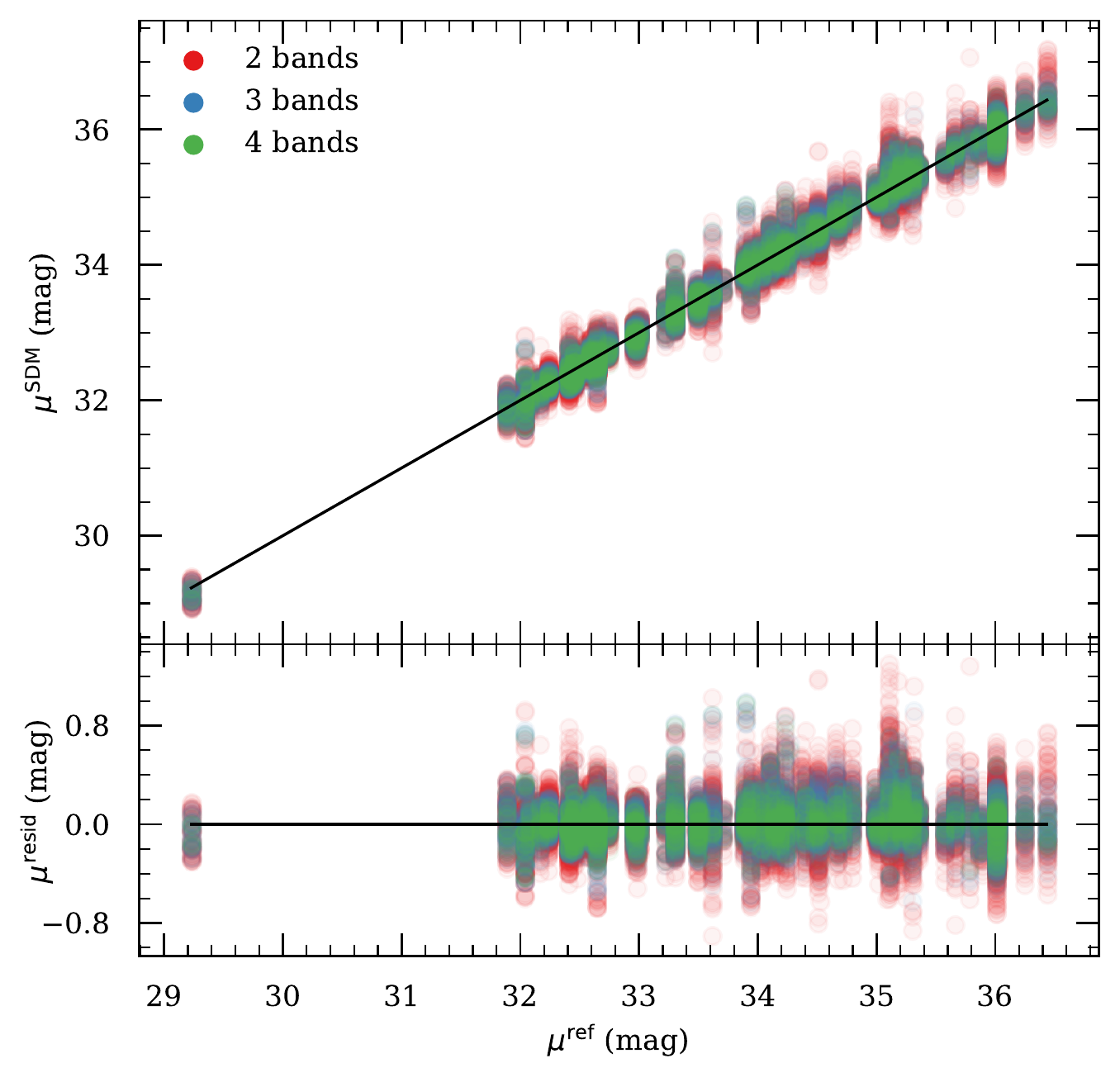}
\caption[Comparison of SDM-derived distance moduli to their \texttt{SNooPy} reference values]{Comparison of SDM-derived distance moduli to their \texttt{SNooPy} reference values, with residuals in the bottom panel. Colours distinguish the number of passbands used for the single photometric epoch in the SDM fit.\label{fig:comparison}}
\end{figure}

The top-level result (see Figure~\ref{fig:comparison}) is encouraging: the median residual across all validation snapshot distances is just $0.008_{-0.124}^{+0.138}$\,mag (16th and 84th percentile differences are reported for scatter in this fashion here and throughout), and as we shall see in the following subsections, an even higher level of performance is realised when we restrict to the more information-rich maskings of our data. By way of comparison, the corresponding result for our ``control'' exercise (where we omit spectroscopically-derived quantities) is $-0.034_{-0.288}^{+0.290}$\,mag.

Despite the notation used above and throughout (which we employ for compactness), we emphasise that the reported scatter values \textit{should not} be confused with uncertainty estimates --- indeed, they are in no way derived from $\mu^\mathrm{SDM}$ error bars. To provide such an uncertainty estimate on a single metric (e.g., $\mu^\mathrm{resid}$) that describes our full set of residuals would be difficult, given the correlations induced by the repetition of spectra and photometry in our validation exercise. Instead, we perform another test where we pick three characteristic points in the photometric evolution of each distinct SN~Ia in our sample: (i) ``Earliest,'' corresponding to the first available photometric epoch for a given object; (ii) ``Nearest to Max,'' for the photometric epoch closest to $t_\mathrm{max}$; and (iii) ``Latest,'' giving the last available photometric epoch. Iterating through each of the 11 distinct passband combinations available for our compilation, we select --- for each distinct SN~Ia in our sample --- the single $\mu^\mathrm{resid}$ value (and its error bar) corresponding to each of these photometric-evolution points, from which we compute the mean and its propagated uncertainty for each. In cases where multiple spectra are available for a given object, we use only the one closest to $t_\mathrm{max}$.

\begin{figure*}
\centering
\includegraphics[width=\textwidth]{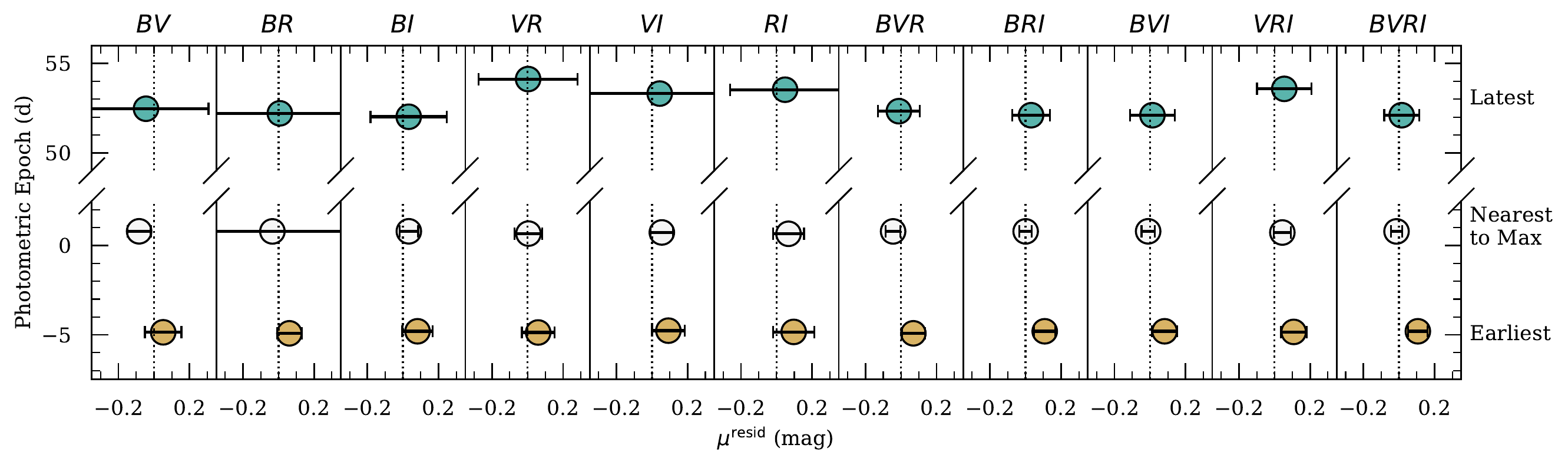}
\caption[Mean residuals as a function of passband combination and representative photometric epoch]{Mean residuals with propagated uncertainties in a grid determined by passband combination and representative photometric epoch. Each point uses a single value for each SN~Ia in our compilation to derive the mean and its propagated uncertainty from SDM-distance error bars, thus avoiding data repetition and the correlations it can induce.\label{fig:stat_err}}
\end{figure*}

These values, shown in Figure~\ref{fig:stat_err}, represent aggregations over nonrepeated data, thus affording a proper uncertainty diagnostic that is free from artifacts introduced by correlation. Across the 33 distinct mean residual values (11 passband combinations $\times$ 3 characteristic photometric-evolution points), 25 are consistent (i.e., within their $1\sigma$ error bars) with zero. Aggregating over the selected photometric evolution points, we find that, given their generally large uncertainty values (median uncertainty: 0.21\,mag) 11/11 ``Latest'' residuals are consistent with zero, while 9/11 ``Nearest to Max'' values (having a median uncertainty of 0.05\,mag) are, and just 5/11 ``Earliest'' values (with a median uncertainty of 0.07\,mag) are. Moreover, all are consistent with zero at the $2\sigma$ level. These results are satisfactory and consistent with our expectations: maximum performance is achieved when the photometric data are near maximum light, but the performance degradation at earlier or later times is not so significant as to mitigate the utility of the SDM.

\subsubsection{Parameter Dependence}

\begin{figure*}
\centering
\includegraphics[width=\textwidth]{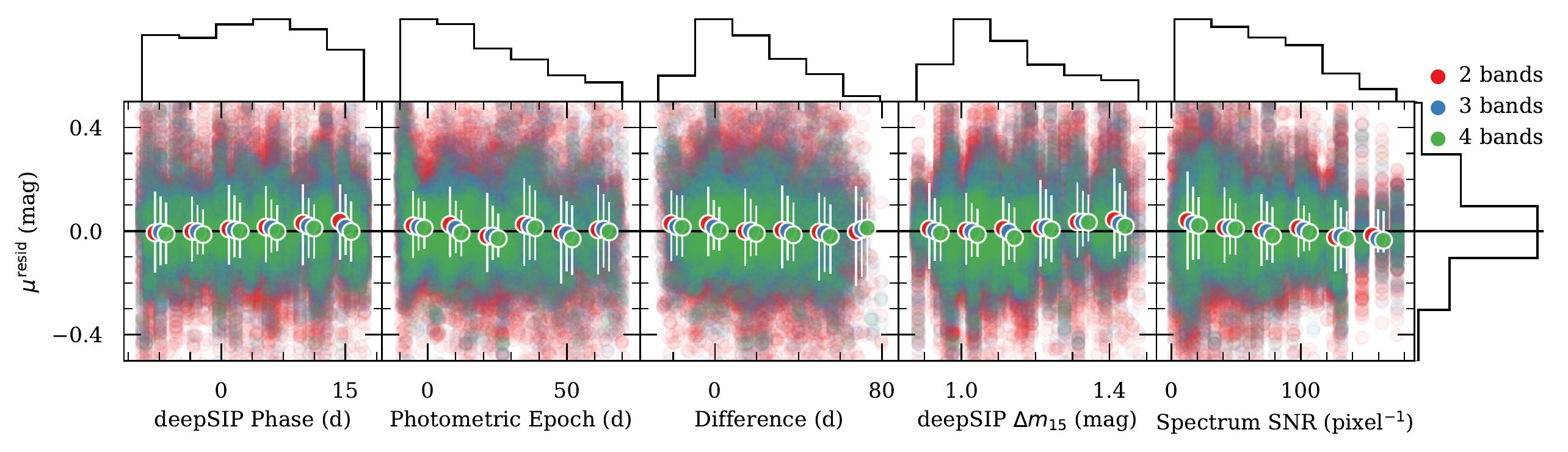}
\caption[Distance-modulus residuals as a function of various photometric and spectroscopic properties]{Distance-modulus residuals (SDM minus \texttt{SNooPy} reference) as a function of (from left to right) \texttt{deepSIP}-predicted phase, rest-frame photometric epoch relative to maximum, rest-frame difference between phase of photometric epoch and phase of spectrum, \texttt{deepSIP}-predicted $\Delta m_{15}$ value, and spectrum SNR (the average pixel size of spectra in our compilation is 1.7 \AA). The distributions of each quantity are projected outside the axes. Data points are colour-coded according to the same scheme as in Figure~\ref{fig:comparison}, and the white error bars signify the median residual (and its 16th and 84th percentile differences) for each fixed-width bin in the upper projections. The error bars --- which are indicative of scatter, not propagated uncertainty --- are slightly horizontally offset as a visual aid to see the number of passbands used in computing them, again denoted by colour.\label{fig:resid-vs-params}}
\end{figure*}

To search for biases in the snapshot distances generated by our validation exercise, we study distance-modulus residuals, $\mu^\mathrm{resid}$, as a function of temporal indicators, luminosity indicators inferred from \texttt{deepSIP} predictions, and spectrum SNRs, each segmented by the number of passbands included in the photometric epoch. The results, conveyed in Figure~\ref{fig:resid-vs-params}, are highly encouraging. We find that the distance-modulus residuals are consistent with zero and show no obvious correlation with \texttt{deepSIP}-predicted phase, rest-frame photometric epoch, rest-frame difference between the phase of the photometric epoch and that of the spectrum, \texttt{deepSIP}-predicted $\Delta m_{15}$, or spectrum SNR. Moreover, we find that both the median absolute residual and scatter decrease as we permit more passbands to be included in the SDM fit (see Table~\ref{tab:sim-results} for a summary of all results, segmented by passband combination).

\begin{table}
{\addtolength{\tabcolsep}{-2pt}
\caption{Validation Results.\label{tab:sim-results}}
\begin{tabular}{crcrcr}
\hline
\hline
Bands & \multicolumn{1}{c}{med$\left(\mu^\mathrm{resid}\right)$} & Bands & \multicolumn{1}{c}{med$\left(\mu^\mathrm{resid}\right)$} & Bands & \multicolumn{1}{c}{med$\left(\mu^\mathrm{resid}\right)$} \\
 & \multicolumn{1}{c}{(mag)} &  & \multicolumn{1}{c}{(mag)} &  & \multicolumn{1}{c}{(mag)} \\
\hline 
 $BI$ &   ${0.031}_{-0.120}^{+0.126}$ &  $BRI$ &   ${0.010}_{-0.092}^{+0.109}$ &  $BVRI$ &  ${-0.003}_{-0.097}^{+0.107}$ \\
 $BR$ &  ${-0.009}_{-0.098}^{+0.107}$ &  $BVI$ &  ${-0.008}_{-0.105}^{+0.109}$ &         &                               \\
 $BV$ &  ${-0.075}_{-0.156}^{+0.143}$ &  $BVR$ &  ${-0.024}_{-0.098}^{+0.107}$ &         &                               \\
 $RI$ &   ${0.053}_{-0.182}^{+0.177}$ &  $VRI$ &   ${0.056}_{-0.115}^{+0.127}$ &         &                               \\
 $VI$ &   ${0.049}_{-0.142}^{+0.145}$ &        &                               &         &                               \\
 $VR$ &   ${0.053}_{-0.158}^{+0.149}$ &        &                               &         &                               \\
 \hline
 2 & ${0.012}_{-0.143}^{+0.156}$ &3 & ${0.007}_{-0.107}^{+0.120}$ &4 & ${-0.003}_{-0.097}^{+0.107}$ \\
 \hline
 \multicolumn{6}{p{\columnwidth}}{\textbf{Note:} The last row shows the results segmented by the number of passbands, instead of the specific combination. Following the convention of this paper, the median values are reported with 16th and 84th percentile differences to show the scale of the scatter.}
\end{tabular}}
\end{table}

As evidenced in Table~\ref{tab:sim-results}, the median residual for each of the 11 distinct passband combinations overlap within their scatter (and are all consistent with zero), but interestingly, the residuals for SDM fits using the $BR$ passband combination perform markedly better than all other two-passband fits (having a median $\sim 3$--8 times closer to zero), and better than all three-passband fits as well (except for $BVI$). The scatter for $BR$ fits is competitive with that for $BVRI$ (and \emph{much} tighter than all other two-passband combinations except $BI$), but the latter outperforms the median residual of the former by a factor of $\sim 3$. Though the origin of the relatively high quality of $BR$ (and to a lesser extent, $BI$) SDM fits remains somewhat unclear, the fact remains that, at least for our data compilation, distances can be estimated to a very satisfactory degree of certainty using just one optical spectrum and two contemporaneous photometric points in distinct passbands (and the relative quality increases as more passbands are added). Moreover, given the relative scale of all scatter values reported in Table~\ref{tab:sim-results} relative to their corresponding median residual values, the discussion in this paragraph is, perhaps, nearing the limit of being overly detailed. We emphasise the main takeaway: that all median residual values are consistent with zero, given the scale of the observed scatter.

\subsubsection{Fit Quality}

\begin{figure*}
\centering
\includegraphics[width=\textwidth]{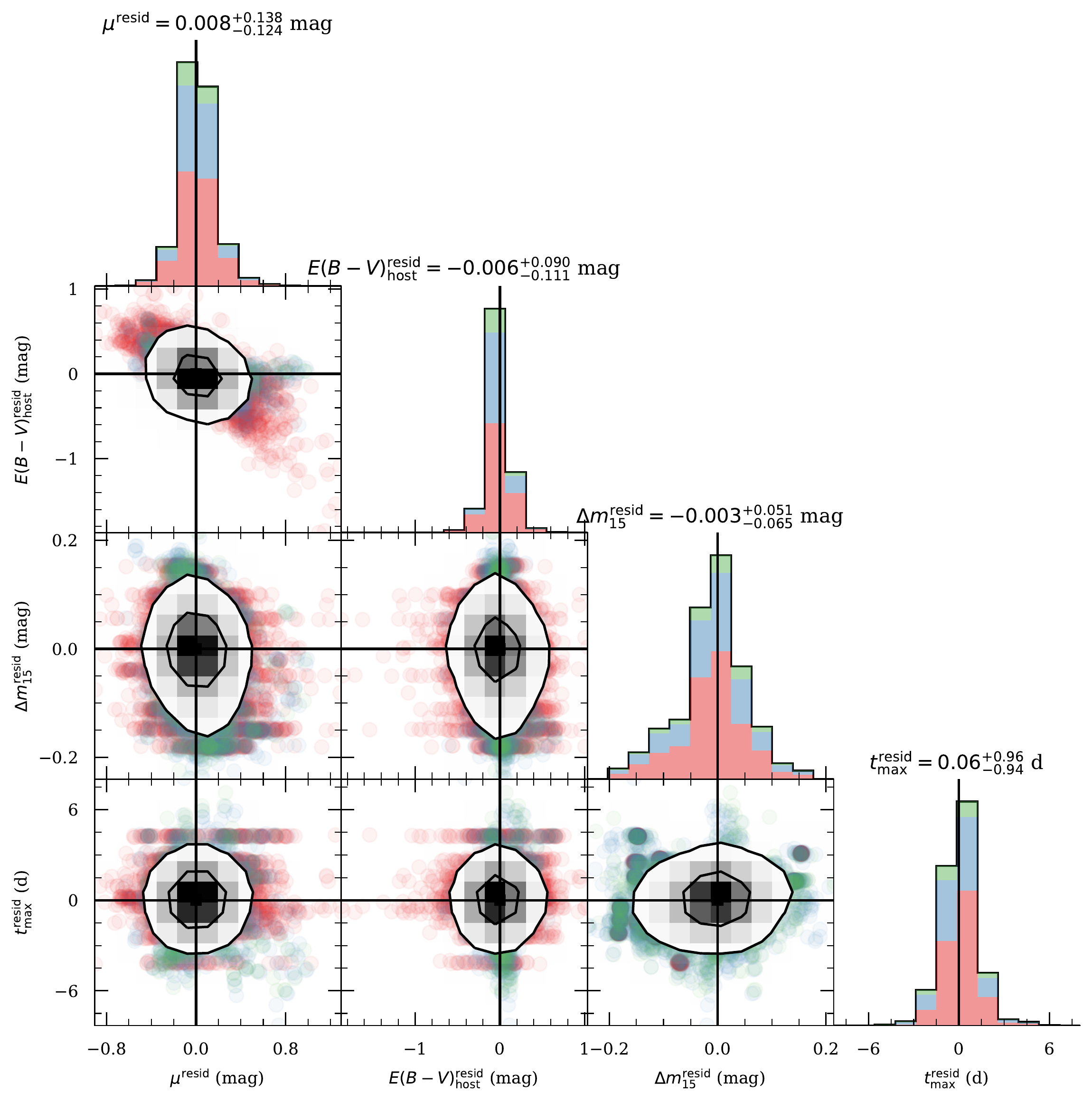}
\caption[One- and two-dimensional projections of parameter residuals]{One- and two-dimensional projections of $E(B-V)$ model parameter residuals between SDM fits and \texttt{SNooPy} reference values \citep[made with the \texttt{corner} package;][]{corner}. Data points and stacked histograms are colour-coded according to the same scheme as in Figure~\ref{fig:comparison}, and smoothed $1\sigma$ and $2\sigma$ contours are given in black. For each set of residuals, the median value and 16th and 84th percentile differences are labeled (the latter as an indicator of scatter, not propagated uncertainty), and vertical and horizontal lines are used to locate the expected zero-residual location.\label{fig:resid-corner}}
\end{figure*}

Next, we investigate the quality of SDM fits by computing the residuals between realised model parameters in such fits and the corresponding \texttt{SNooPy} reference values for a given SN~Ia. We present the one- and two-dimensional distributions of these residuals in Figure~\ref{fig:resid-corner}, and we note that they are best grouped into two categories: those determined largely by \texttt{deepSIP} from spectra (e.g., $t_\mathrm{max}$ and $\Delta m_{15}$) and those derived via the MCMC fit [e.g., $\mu$ and $E(B-V)_\mathrm{host}$].

The former (i.e., $t_\mathrm{max}$ and $\Delta m_{15}$ residuals) are mostly --- but not exclusively, as we shall shortly discuss --- a measure of the quality of \texttt{deepSIP} predictions. From the one-dimensional distributions in Figure~\ref{fig:resid-corner}, we can see that \texttt{deepSIP}-predicted $\Delta m_{15}$ values in aggregate fall within $\sim 0.005$ mag of the corresponding values from a fit to all photometry, and \texttt{deepSIP}-inferred $t_\mathrm{max}$ values to within $\sim 0.05$ d. Perhaps unsurprisingly, these metrics (derived from the final MCMC fit described in Section~\ref{sec:method}) represent a modest improvement over what is obtained by doing the preliminary least-squares fit. More significantly, the fact that we perform an MCMC fit is why the aforementioned metrics are not a perfect measure of the quality of \texttt{deepSIP} predictions --- we allow our estimates for $t_\mathrm{max}$ and $\Delta m_{15}$, initially derived from an optical spectrum via \texttt{deepSIP}, to be updated in light of additional evidence: the photometric data. The fact that this procedure leads to superior agreement is a very promising result indeed; \texttt{deepSIP} predictions provide an excellent starting point for $t_\mathrm{max}$ and $\Delta m_{15}$, but both are generally even better fit when photometric information is taken into account. Moreover, it is precisely because of this that the ``clumpiness'' in the $t^\mathrm{resid}_\mathrm{max}-\Delta m^\mathrm{resid}_{15}$ distribution in Figure~\ref{fig:resid-corner} gets modestly blurred out. This inherent ``clumpiness'' is expected owing to the fact that $t_\mathrm{max}$ and $\Delta m_{15}$ are properties of a specific SN~Ia, not a specific spectrum: many distinct points corresponding to different spectra, photometric epochs, and passband combinations should map to exactly the same point in the $t^\mathrm{resid}_\mathrm{max}-\Delta m^\mathrm{resid}_{15}$ distribution. We also note that the scatters in residuals ($\sim 0.06$\,mag for $\Delta m_{15}$ and $\sim 1$\,d for $t_\mathrm{max}$) are broadly consistent with the findings of S20 when they studied the quality of \texttt{deepSIP} predictions.

Moving now to the latter set of residuals [i.e., $\mu$ and $E(B-V)_\mathrm{host}$], we first remark on the quality of the fits: in aggregate, $\mu$ is recovered to $\lesssim 0.01$\,mag with a scatter of $\sim 0.13$\,mag while $E(B - V)_\mathrm{host}$ is recovered even more closely and tightly. As our focus in this work is on \emph{distances}, we limit the following discussion to $\mu$ except for where $E(B - V)_\mathrm{host}$ has an impact. Looking at the left-most bottom two panels in Figure~\ref{fig:resid-corner}, we are encouraged to see nearly negligible dependence of the $\mu$ residuals on those for $t_\mathrm{max}$ or $\Delta m_{15}$. Indeed, the colour scale implies that the number of passbands used in the SDM fit has a much bigger impact on the quality of $\mu$ predictions, with 2-band fits (signified by red dots) visibly protruding from the horizontal edges of the $2\sigma$ contours, thereby broadening the distribution of distance-modulus residuals. The full, 4-band $BVRI$ fits (green dots), when visible, only ``bleed'' out from the top and bottom of the contours, thus maintaining the narrow subdistribution that is seen in the one-dimensional $\mu^\mathrm{resid}$ distribution at the top of Figure~\ref{fig:resid-corner}. There is some evidence for an inverse correlation between $\mu$ residuals and those for $E(B - V)_\mathrm{host}$, but this is unsurprising given the form of Equation~\ref{eqn:ebv}: since the sum of $\mu + E(B - V)_\mathrm{host}$ is in part responsible for the observed magnitude, an overestimate by one can be compensated by an underestimate of the other. Regardless, the effect of this degeneracy is most pronounced for the 2-band fits which, as shown in Table~\ref{tab:sim-results}, underperform the 3-band and 4-band fits in most regards.

\section{Discussion}
\label{sec:discussion}

\subsection{Summary}

As we have shown in Section~\ref{sec:method}, it is possible to generate robust ``snapshot'' distance estimates to SNe~Ia from a very limited observing expenditure --- one optical spectrum and one epoch of multipassband photometry per object are sufficient. The optical spectrum, via \texttt{deepSIP}, delivers the \emph{intrinsic} photometric evolution (parameterised in our case by $t_\mathrm{max}$ and $\Delta m_{15}$), and the epoch of photometry provides a sampling of this evolution, but in the \emph{observer} frame. The discrepancy between the former and the latter is due \emph{mostly} to the inverse-square law of light (i.e., because of the distance to the object), and the remainder can be appropriately modeled and accounted for using the colour information provided by the photometry.

To test the efficacy of our method, we assemble a compilation of SN~Ia spectra with corresponding well-observed light curves from a larger set that was uniformly prepared by S20. Using this compilation, we generate $> 30,000$ snapshot distances by providing every possible combination of one spectrum and one epoch of $2+$ passband photometry that fall within minimally restrictive temporal bounds. We then compare these validation snapshot distances to the corresponding distances obtained from a light-curve fit to all available photometry (taken reference values), and use the residuals as an overall probe of our method's ability to reproduce the latter, \texttt{SNooPy} reference values.

To this end, our method performs very well, with a median residual between \emph{all} snapshot distance moduli and their \texttt{SNooPy} reference counterparts of just $0.008_{-0.124}^{+0.138}$\,mag. Performance roughly at this level is maintained over a wide range of rest-frame spectral ($-10$--18\,d) and photometric ($-10$--70\,d) phases, as well as $\Delta m_{15}$ values (0.85--1.5\,mag) and spectrum SNRs. Indeed, our investigations reveal that a much stronger determinant of performance is the number of passbands available in the single epoch of photometry --- in aggregate, the median absolute residual and scatter decrease as we supply more contemporaneous photometric points in distinct passbands, reaching a level of $-0.003_{-0.097}^{+0.107}$\,mag (a scatter in distance of just $\sim 5\%$ relative to the ``true'' values) when each epoch includes information in $BVRI$. This trend follows our intuition that SDM fits should be better constrained and hence of higher quality as more data are provided. Interestingly, however, one 2-passband combination ($BR$) and three 3-passband combinations ($BRI$, $BVI$, and $BVR$) have a similarly small degree of scatter, but none produces a median residual quite as close to zero.

\subsection{Variations of the Snapshot Distance Method}

Returning to the aforementioned trend (of increasing quality as more data are provided), we can investigate the extent to which adding data along the temporal dimension --- as opposed to the wavelength dimension, which is accomplished by adding more contemporaneous passbands --- reduces scatter. Such supplementation of temporal data can be accomplished either by providing additional spectra or additional epochs of multiband photometry, and the results can help us to identify which component of our method has the largest leverage in reducing the observed scatter in $\mu^\mathrm{resid}$. We therefore study both as follows.

\subsubsection{Additional Spectra}

The former is straightforward to implement. We simply repeat the validation exercise described in Section~\ref{ssec:simulation}, except that instead of iterating over all spectra, we step through the 50 SNe~Ia having at least two spectra in our compilation and set $t_\mathrm{max}$ and $\Delta m_{15}$ for each object as the mean of the \texttt{deepSIP}-inferred values from all of the available spectra. Uncertainties are derived through error propagation.

The top-level result is $\mu^\mathrm{resid} = 0.013_{-0.111}^{+0.129}$\,mag, consistent with the corresponding metric from our original validation exercise, albeit with slightly reduced scatter. A similar trend is realised when we look at the $BR$ ($\mu^\mathrm{resid} = {-0.008}_{-0.084}^{+0.090}$\,mag) and $BVRI$ ($\mu^\mathrm{resid} = {0.002}_{-0.071}^{+0.085}$\,mag) subsets. We find a ``sweet spot'' of $\sim 3$ spectra per object where the scatter is further reduced, and although one might expect a continuing trend of reduction as more spectra are provided, we do not see this in our \emph{low-number-statistics} data for $> 4$ spectra per object. In any case (and independent of the number of spectra provided), the improved metrics noted above only modestly outperform the results of our original validation exercise --- spectra appear \emph{not} to be the origin of most of the observed scatter. We take this to be an indication of the quality of \texttt{deepSIP} predictions: one spectrum, truly, is sufficient to robustly estimate $t_\mathrm{max}$ and $\Delta m_{15}$.

\subsubsection{Additional Photometry}

Testing the latter --- i.e., incorporating additional epochs of photometry --- is much more computationally expensive. With $(2,450 \mathrm{\:total\:epochs})/(97 \mathrm{\: SNe\:Ia}) \approx 25$ epochs of multiband photometry per object, on average, the task of handling all possible combinations of two epochs grows by a factor of 12 relative to the one-epoch case, and by a factor of 92 for the three-epoch case. As our primary validation exercise already takes $\sim 26$\,hr to run on a modern 20-core server, it is hard to justify an even larger expenditure for this investigation, and even if we did, an exhaustive search over all possible combinations is simply intractable; e.g., there are $> 5\times 10^6$ ways to select a sample of 12 from 25. We therefore perform another validation exercise, but instead of selecting all possible combinations, $N {\mathrm C} n$, where $N$ is the number of photometric epochs available for a given SN~Ia and $1 \leq n < N$ is the subset size (this is the very expensive part), we select subsets using a simple ``dilution'' factor, $\phi$. Specifically, we perform an identical exercise to that described in Section~\ref{ssec:simulation}, except that instead of masking all but one epoch, we mask all but every $\phi$th epoch for $\phi = 2, 3, \cdots, 12$ (the special case of $\phi = 1$ corresponds to a fit to all available photometry, which we refer to as ``\texttt{SNooPy} reference'' throughout).

Unsurprisingly, the best results are found with $\phi = 2$ (corresponding to the densest temporal sampling, yielding $\mu^\mathrm{resid} = {0.011}_{-0.071}^{+0.075}$\,mag), and particularly for the subset provided with simultaneous\footnote{We consider photometric points within $\pm 0.001$\,d of one another to be simultaneous.} $BVRI$ information in each epoch ($\mu^\mathrm{resid} = {0.005}_{-0.027}^{+0.021}$\,mag). In the case of $\phi = 12$ (i.e., $\sim 2$ photometric epochs per SN~Ia, on average) the corresponding metrics grow (in scatter) to ${0.010}_{-0.097}^{+0.101}$\,mag and ${0.003}_{-0.079}^{+0.067}$\,mag, respectively. In comparing against ``control'' exercises (where we do not provide spectroscopically-derived quantities), we find comparable levels of performance up to $\phi = 4$ and then a growing trend of SDM outperforming the control with increasing $\phi$ (i.e., as the photometric coverage becomes more sparse). This suggests that above a certain threshold of photometric coverage, the SDM is consistent with, but not necessarily superior to, a conventional light-curve fit, but below this threshold, it offers significantly improved prediction power (as is our expectation).

As a result of this, we can identify two ``levers'' that wield significant influence over the scatter in $\mu^\mathrm{resid}$: (i) the number of simultaneous passbands provided per epoch (i.e., the wavelength-space extent of the SN~Ia spectral energy distribution sampled at a specific instant; as we have concluded above, more is better), and (ii) the number of distinct multiband epochs of photometry available for the fit (i.e., the temporal-space extent of the SN~Ia spectral energy distribution evolution, with sampling provided by the photometric observations; more is better). Neither appears to be decisively stronger than the other; e.g., the size of the scatter reduction in taking $\phi = 12 \to 2$ in the global set is roughly the same as that harnessed in holding fixed $\phi = 12$ and going from the global set to the $BVRI$ subset. This suggests that sparser temporal coverage can be compensated for by providing more extensive wavelength coverage (i.e., supplying more passbands per epoch). Though it is beyond the scope of this study, it would be interesting to examine how well these conclusions are rederived by a more extensive (and necessarily, expensive) validation exercise that makes fewer simplifying assumptions than we have invoked here.

\subsection{Applications and Future Work}

There are, of course, many further variations that one could explore with regard to our method (e.g., the photometric contemporaneity requirement could be relaxed). Nevertheless, the studies presented herein demonstrate that the distance to an SN~Ia can be robustly estimated from just one night's worth of observations, and that when more data are available, the estimates improve in quality until reaching a level of consistency with conventional light-curve fits. Thus, in the coming era of wide-field, large-scale surveys, our snapshot distance method will ensure maximal scientific utility from the hundreds of thousands of SNe~Ia that will be discovered, but which may not be sufficiently well monitored to derive reliable distances by conventional means. Moreover, our method holds the prospect of ``unlocking'' a significant number of otherwise unusable observations that currently exist. As a case study, we present (in a companion paper) our use of the SDM to estimate the distances to $>100$ sparsely observed SNe~Ia which, when combined with a literature sample, deliver cutting-edge constraints on the cosmological parameter combination, $f\sigma_8$ \citep{StahlPV}.

\section*{Acknowledgements}

We thank an anonymous referee for very thorough suggestions that led to significant improvements in this paper. B.E.S. and T.d.J. thank Marc J. Staley and Gary \& Cynthia Bengier (respectively) for generously providing fellowship funding. B.E.S. also thanks Keto Zhang for a series of conversations that proved valuable in the evolution of this work. A.V.F. has been generously supported by the TABASGO Foundation, the Christopher R. Redlich Fund, and the U.C. Berkeley Miller Institute for Basic Research in Science (where he is a Senior Miller Fellow).
This research used the Savio computational cluster resource provided by the Berkeley Research Computing program at U.C. Berkeley (supported by the Chancellor, Vice Chancellor for Research, and Chief Information Officer).

\section*{Data Availability}

All data used herein are publicly available through the references described in Section~\ref{ssec:data}.




\bibliographystyle{mnras}
\bibliography{bib}






\bsp	
\label{lastpage}
\end{document}